\documentclass{aa}

\usepackage[]{graphics}
\usepackage[]{psfig}

\begin{document}

\thesaurus{}

\title{$\omega$ Centauri -- a former nucleus of a dissolved dwarf galaxy? New
evidence from Str\"omgren photometry}

\author {Michael Hilker \inst{1} \and Tom Richtler \inst{2}
}

\offprints {M.~Hilker}
\mail{mhilker@astro.puc.cl}

\institute{
Departamento de Astronom\'\i a y Astrof\'\i sica, P.~Universidad Cat\'olica,
Casilla 104, Santiago 22, Chile (mhilker@astro.puc.cl)
\and
Departamento de F\'\i sica, Universidad de Concepci\'on, Casilla 160-C,
Concepci\'on, Chile (tom@coma.cfm.udec.cl)
}

\date {Received 26 June 2000 / Accepted 28 August 2000}

\titlerunning{$\omega$ Centauri -- a nucleus of a dissolved dwarf galaxy}

\authorrunning{M.~Hilker \& T.~Richtler}
\maketitle

\begin{abstract}

CCD $vby$ Str\"omgren photometry of a statistically complete sample of
red giants and stars in the main sequence turn-off region in $\omega$
Centauri has been used to analyse the apparently complex star formation 
history of this cluster. From the location of stars in the $(b-y),m_1$ 
diagram metallicities have been determined. These have been used to estimate 
ages of different sub-populations in the color-magnitude diagram and to 
investigate their spatial distributions. We can confirm several earlier 
findings. The dominating metal-poor population around $-1.7$ dex is the 
oldest population found. More metal-rich stars between [Fe/H]=$-1.5$ and 
$-1.0$ dex tend to be 1-3 Gyr younger. These stars are more concentrated 
towards the cluster center than the metal-poor ones. The most-metal rich 
stars around $-0.7$ dex might be up to 6 Gyr younger than the oldest 
population. They are asymmetrically distributed around the center with an 
excess of stars towards the South.

We argue that the Str\"omgren metallicity in terms of element abundances
has another meaning than in other globular clusters. From a comparison
with spectroscopic element abundances, we find the best correlation with
the sum C+N. The high Str\"omgren metallicities, if interpreted by strong 
CN-bands, result from progressively higher N and perhaps C abundances in 
comparison to iron. The large scatter of Str\"omgren abundances may come 
from a variety of evolutionary effects, including C-depletion and 
N-enrichment. We see an enrichment already among the metal-poor population, 
which is difficult to explain by self-enrichment alone.

An attractive speculation (done before) is that $\omega$ Cen was the nucleus 
of a dwarf galaxy. We propose a scenario in which $\omega$ Cen experienced  
mass inflow over a long period of time, until the gas content of its host 
galaxy was so low that star formation in $\omega$ Cen stopped, or 
alternatively the gas was stripped off during its infall in the Milky Way 
potential.  This mass inflow could have occurred in a clumpy and discontinuous 
manner, explaining the second peak of metallicities, the abundance pattern, 
and the asymmetric spatial distribution of the most metal-rich population. 
Moreover, it explains the kinematic differences found between metal-poor and 
metal-rich stars.


\end{abstract}


\section{Introduction}

Since decades $\omega$ Centauri has been the target of many investigations
by photometric and spectroscopic means. It is known as an extraordinary object
among the star clusters of our Galaxy. $\omega$ Cen is not only the most
massive and one among the most flattened Galactic globular clusters (Meylan
\cite{meyl87}; White \& Shawl \cite{whit87b}), but also shows strong variations
in nearly all element abundances investigated so far (e.g. Norris \& Da Costa
\cite{norr95} and references therein). This  is reflected
by the intrinsic broad scatter of the red giant branch (already noticed in
the 70s by Cannon \& Stobie \cite{cann}) that cannot be explained by internal
reddening only (e.g. Norris \& Bessell \cite{norr75}). Most recently,
Pancino et al. (\cite{panc}) confirmed the existence of a very metal-rich
population in $\omega$ Cen whose red giant branch (RBG) is well separated
from the bulk of the RGB stars.

\begin{table*}
\caption[]{\label{log} Observation log of $\omega$ Centauri fields}
\begin{flushleft}
\begin{tabular}{ccccrrrc}
\hline\noalign{\smallskip}
Id. & $\alpha_{2000}$ & $\delta_{2000}$ & Date & \multicolumn{3}{c}{Exposure
time [s]} & seeing \\
 & & & & $y$ & $b$ & $v$ & \\
\noalign{\smallskip}
\hline\noalign{\smallskip}
1 & 13:26:57.9 & $-$47:33:09 & 1993 May 14 & 30,60      & 60,120     &
120,240      & $1\farcs4$--$1\farcs7$ \\
2 & 13:26:59.7 & $-$47:38:50 & 1993 May 12 & 30,360,240 & 60,660,600 &
120,960,960  & $2\farcs0$--$2\farcs5$ \\
  &            &             & 1993 May 14 & 30         & 60,240     &
120,480      & $1\farcs4$--$1\farcs6$ \\
3 & 13:27:00.1 & $-$47:41:52 & 1993 May 14 & 30,300,300 & 60,600,600 &
120,900,900  & $1\farcs4$--$1\farcs9$ \\
4 & 13:27:00.1 & $-$47:46:02 & 1993 May 14 & 30,300,300 & 60,600,600 &
120,960,1020 & $1\farcs4$--$1\farcs7$ \\
5 & 13:27:26.3 & $-$47:28:54 & 1993 May 13 & 30,300,360 & 60,600,720 &
120,960,1080 & $1\farcs1$--$1\farcs4$ \\
6 & 13:27:55.6 & $-$47:28:01 & 1993 May 13 & 30,300,360 & 60,600,660 &
120,900,1020 & $1\farcs3$--$1\farcs5$ \\
7 & 13:28:21.1 & $-$47:27:52 & 1993 May 13 & 30,300,360 & 60,600,660 &
120,900,1020 & $1\farcs2$--$1\farcs4$ \\
8 & 13:28:49.4 & $-$47:27:52 & 1993 May 13 & 30,300     & 60,600     &
120,900      & $1\farcs3$--$1\farcs5$ \\[1.5mm]
A & 13:28:01.1 & $-$47:28:32 & 1995 Apr 21 & 30,360     & 60,360     &
120,720      & $1\farcs3$--$1\farcs6$ \\
  &            &             & 1995 Apr 22 & 30,120     & 60,210     &
120,420      & $1\farcs4$--$1\farcs7$ \\
B & 13:27:48.0 & $-$47:24:06 & 1995 Apr 21 & 30,300,240 & 60,480,420 &
120,780,720  & $1\farcs2$--$1\farcs6$ \\
C & 13:27:17.0 & $-$47:22:26 & 1995 Apr 21 & 30,240,180 & 60,420,360 &
120,720,840  & $1\farcs1$--$1\farcs3$ \\
D & 13:26:50.0 & $-$47:19:06 & 1995 Apr 21 & 30,90      & 60,180     &
120,420      & $1\farcs1$--$1\farcs2$ \\
E & 13:26:42.0 & $-$47:24:06 & 1995 Apr 21 & 30,180     & 60,360     &
120,600      & $1\farcs1$--$1\farcs2$ \\
  &            &             & 1995 Apr 22 & 30,180,180 & 60,300,300 &
120,600,600  & $1\farcs2$--$1\farcs4$ \\
F & 13:26:15.0 & $-$47:26:56 & 1995 Apr 22 & 30,150,150 & 60,300,270 &
120,600,600  & $1\farcs2$--$1\farcs4$ \\
G & 13:25:45.9 & $-$47:26:56 & 1995 Apr 22 & 30,300,300 & 60,540,540 &
120,900,900  & $1\farcs4$--$1\farcs6$ \\
H & 13:25:15.9 & $-$47:26:56 & 1995 Apr 22 & 30,240,180 & 60,480,300 &
120,840,600  & $1\farcs3$--$1\farcs6$ \\
I & 13:24:45.8 & $-$47:26:57 & 1995 Apr 24 & 30,180     & 60,330     &
120,660      & $1\farcs2$--$1\farcs3$ \\
0 & 13:26:50.0 & $-$47:28:33 & 1995 Apr 24 & 30         & 60         &
120          & $1\farcs1$--$1\farcs2$ \\
\noalign{\smallskip}
\hline
\end{tabular}
\end{flushleft}
\end{table*}

The abundance variations in $\omega$ Cen point to a more complicated
star formation history than that for other globular clusters which
contain a homogeneous  stellar population.
Whereas the CNO variations might be explained by evolutionary mixing effects
in the stellar atmosphere as well as by mixing in the protocloud (e.g.
Bessell \& Norris \cite{bess76}), the iron abundance
variations need another explanation (e.g. Vanture et al. \cite{vant}, Norris \&
Da Costa \cite{norr95}). The metallicity distribution of the stars in $\omega$
Cen, as derived from Calcium abundance measurements of more than 500 red
giants (Norris et al. \cite{norr96}), indicates a bimodality with a distinct
peak at [Ca/H] $\simeq -1.4$ dex and a shallower peak at -0.9 dex.
These authors also quote statistical evidence that the  more metal-rich stars
are stronger concentrated towards the cluster center, as one may expect from
a dissipative settling of the gas in a self-enrichment process.
Similar results were obtained by Suntzeff \& Kraft (\cite{sunt96}) who
measured Calcium abundances of $\simeq$ 380 stars and also
found a peak at [Fe/H] $= -1.7$ dex and a broad tail to higher metallicities.
Their data did not support a bimodal metallicity distribution, but again a
weak radial metallicity gradient.
Both groups favoured an extended period of star formation connected with
self-enrichment as the interpretation of their data.
A dynamical analysis of 400 stars in the Norris et al. (\cite{norr96}) sample
revealed a rotation of the metal poor component, whereas the metal rich one is
not rotating (Norris et al. \cite{norr97}). This may be compatible with a
merger of two globular clusters with different masses, as shown by the model
calculations of Makino et al. (\cite{maki}). However, then one would expect
a metallicity distribution that is peaked at two distinct values, which is
in contradiction to the continuous distribution shown by the authors.

A recent spectroscopic investigation by Smith et al. (\cite{smith00}) shows
no strong signatures of enrichment by SNe Ia, but a strong increase of
s-process heavy elements with iron abundances, as has been found by Norris 
\& Da Costa (\cite{norr95}) and other studies as well. This can pose a 
problem, since these elements are believed to come from AGB stars (see Smith 
et al. \cite{smith00} for a discussion and references), and SNe Ia probably
also have intermediate-age progenitors (e.g. McMillan \& Ciardullo
(\cite{mcmi96}).

Because of its peculiar properties, more and more authors raised the question,
whether $\omega$ Cen can actually be regarded as a ``true'' globular cluster,
or whether it is more likely the stripped nucleus of a dwarf galaxy that has
been accreted by the Milky Way (Majewski et al. \cite{maje99b}, Lee et al.
\cite{leey}, Hughes \& Wallerstein \cite{hugh}).
This scenario has also been proposed for M54, one of the most massive globular
clusters. It is a candidate for the nucleus of the Sagittarius dwarf galaxy 
(e.g. Bassino \& Muzzio \cite{bass95}). Three more globular clusters might
have belonged to Sagittarius (Da Costa \& Armandroff \cite{daco95}), now
be added to the Milky Way globular cluster system.

Similarly, the globular clusters NGC 362 and NGC 6779 might have been
associated with the former host galaxy of $\omega$ Cen because of their
similar, strong retrograde orbits (Dinescu et al. \cite{dine99b}).

An origin of $\omega$ Cen within a dwarf galaxy outside the Milky Way could 
provide a natural explanation for its unique properties. Local Group dwarf 
spheroidals are known to have complex star formation histories (e.g. Grebel
\cite{greb97}). One can imagine that this could also be true for the nuclei 
of dwarf galaxies which either experienced a extended period of star formation 
from enriched gas retained in the galaxy potential, or even were built from a
merger of two clusters that spiraled into the center of the dwarf galaxy 
(Miller et al. \cite{mill98}). Medium-resolution spectroscopy of 25 nuclei in 
dwarf ellipicals of the Fornax cluster indicated that all these nuclei have a
metal-rich component and some even may be sites for recent star formation
(Held \& Mould \cite{held}).

However, whether the metallicity spread in $\omega$ Cen also is accompanied by
an age spread awaits verification. In most investigations age effects cannot be 
disentangled from metallicity effects in the turn-off and giant branch region.
Hughes \& Wallerstein (\cite{hugh}) observed one field in $\omega$ Cen with
Str\"omgren $vby$ photometry which is more metallicity sensitive than broad
band photometry. They report an age spread of at least 3 Gyr for stars between
$-2.2 <$ [Fe/H] $<-0.5$ dex (the metal-richest stars also being the youngest).
However, a study of $\omega$ Cen in the Washington system combined with the 
DDO51 filter, which is similarly sensitive to metallicity as the Str\"omgren
system, could not confirm this result (e.g. Majewski et al. \cite{maje99b}).
Further photometric and spectroscopic analysis with a high metallicity and age
resolution are needed to fully understand the complex formation and enrichment
history of this outstanding object.

In this paper we present CCD Str\"omgren $vby$ photometry of about 810
arcmin$^2$ area within $24\arcmin$ from the cluster center.
Str\"omgren photometry has been proven to be a very useful metallicity
indicator for globular cluster giants and subgiants (e.g. Richter et al.
\cite{richp}, Hilker \cite{hilk00a}, Grebel \& Richtler \cite{greb92},
Richtler \cite{rich89}). The location of
late type stars in the Str\"omgren $(b-y),m_1$ diagram is correlated with
their metallicities, especially with their iron and CN abundances.
A new Str\"omgren metallicity calibration for red giants is presented by
Hilker ({\cite{hilk00a}).
With our observations we are able to determine Str\"omgren metallicities 
of a statistically complete number of red giants in $\omega$ Cen and study 
their spatial distribution.

\begin{figure}
\psfig{figure=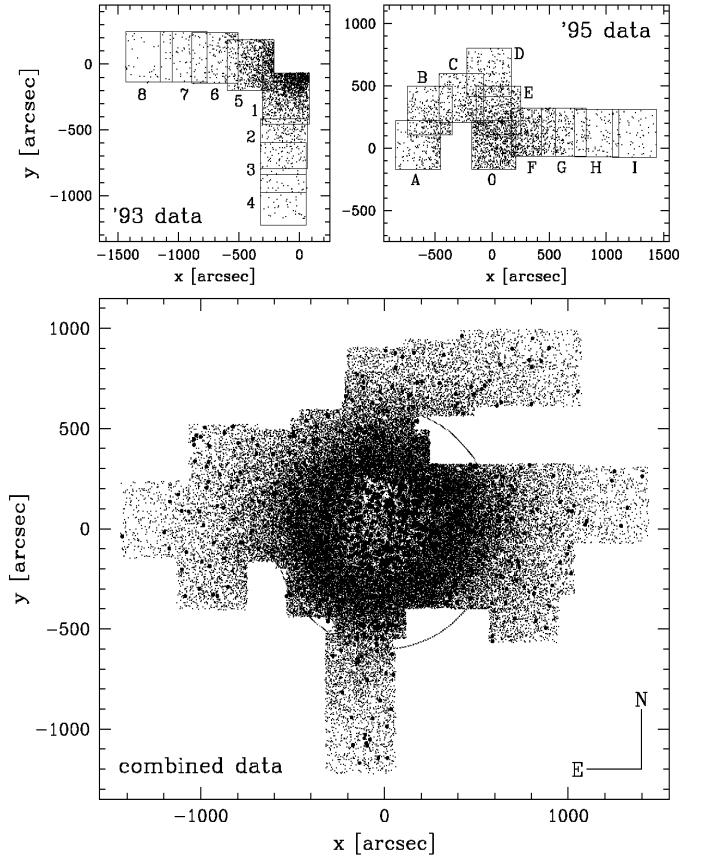,height=11.3cm,width=8.6cm
,bbllx=14mm,bblly=65mm,bburx=152mm,bbury=231mm}
\vspace{0.4cm}
\caption{\label{pos} The lower panel shows the position plot of all observed
fields in $\omega$ Centauri. All stars with a $V$ magnitude brighter than 19.0
mag and a photometric error less than 0.1 mag have been plotted. Bold dots
indicate the position of all stars with determined metallicities (see
Fig.~\ref{pred}). The indicated circle (at 10 arcmin radius) is a selection 
criterium for the analysis of the angular distribution of the RGB stars.
The upper panel shows the positions and nomenclature (see Table \ref{log}) of
the long exposures in the '93 run (left) and the '95 run
(right). Only stars with $V < 16.0$ have been plotted
}
\end{figure}

In Sect.~2 we present the observations and data reduction. Sect.~3 presents
color magnitude diagrams and selection criteria that have been used to
determine the metallicity distribution from the $(b-y),m_1$ diagram (Sect.~4).
In Sect.~5 the red giants in $\omega$ Cen have been divided into
sub-populations, whose ages and spatial distribution have been analysed. 
Finally, Sect.~6 suggests a formation history for $\omega$ Cen and Sect.~7 
summarises the results. 

\section{Observations and data reduction}

The observations of $\omega$ Centauri have been performed in two observing
runs in 1993 and 1995 with the Danish 1.54m telescope (direct imaging) at
ESO/La Silla. In both runs, the CCD in use was a Tektronix chip with
1024$\times$1024 pixels. The $f$/8.5 beam of the telescope provides a scale
of $15\farcs7$/mm, and with a pixel size of 24 $\mu$m the total field is
$6\farcm3 \times 6\farcm3$.
In the first run, 11.-15. May 1993, 8 fields have been observed through the
Str\"omgren $vby$ filters. The nights were not always photometric, and the
seeing, measured by the FWHM of stellar images, was in the range
$1\farcs1$--$2\farcs5$ (see Table~\ref{log}).
In the second run, 21.-24. April 1995, 10 fields have been taken. All nights
had photometric conditions, and the seeing varied between
$1\farcs1$--$1\farcs7$.
The observation log of both runs is given in Table \ref{log} and the position
of the fields are illustrated in Fig.~\ref{pos}.
Additionally, 20 fields with short exposures have been observed in order
to cover the spectroscopic sample of 40 red giants (Norris \& Da Costa
\cite{norr95}), which have been included in a new Str\"omgren metallicity
calibration (Hilker \cite{hilk00a}). The observation and data reduction of
these fields is described in Hilker (\cite{hilk00a}).

The CCD frames were processed with standard IRAF routines, instrumental
magnitudes were derived using DAOPHOT II (Stetson \cite{stet87},
\cite{stet92}).
For the comparison with the standard stars, aperture--PSF shifts have been
determined in all fields. The remaining uncertainty of this shift is in the
order of 0.01 mag in all filters. For the '95 run the corrected magnitudes of
the stars belonging to overlapping areas of two adjacent fields agree very
well and have been avaraged for the final photometry file.
For the photometric reference the '95 run have been chosen, since the '93 run
was not continuously photometric. The instrumental magnitudes determined in
this run have been transformed to those of '95 using stars belonging to
overlapping areas.
The calibration of the combined sample was done via standard stars by
J{\o}nch-S{\o}rensen (\cite{jonc93}, \cite{jonc94}) obtained in the '95 run.
The calibration equations and coefficients are given
in Richter et al. (\cite{richp}), in which the Str\"omgren photometry of M55
and M22 is presented. All Str\"omgren colors refer to the photometric system
defined by Olsen (\cite{olse93}).

After the photometric reduction, combining of the data sets, and calibration
of the magnitudes, the average photometric errors for the red giants used for
the metallicity determination are 0.012 mag for $V$, 0.016 mag for $(b-y)$
and 0.025 mag for $m_1$.

%
%
\section{The Color Magnitude Diagram (CMD)}

\begin{figure}
\psfig{figure=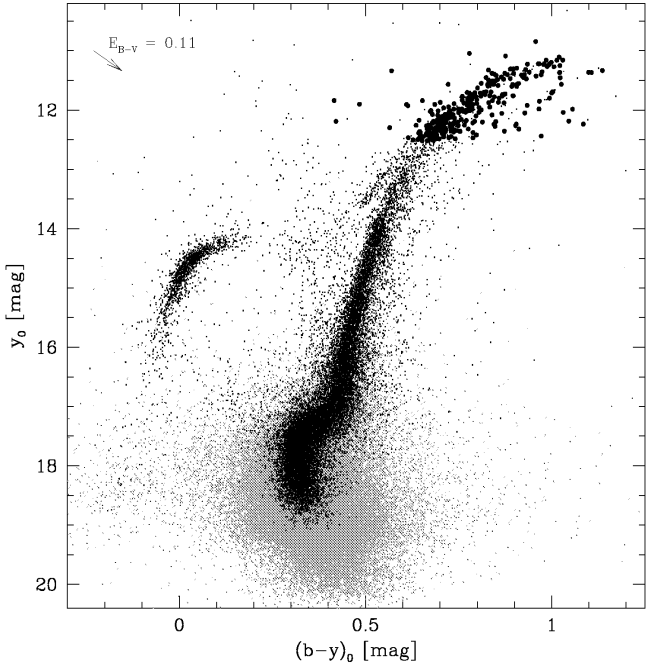,height=8.6cm,width=8.6cm
,bbllx=14mm,bblly=65mm,bburx=188mm,bbury=231mm}
\vspace{0.4cm}
\caption{\label{cmd1}Color magnitude diagram of all observed stars in $\omega$
Cen with a photometric error of less than 0.1 mag (grey) and of less than 
0.03 mag (black) in $y$ and $b$.
The broad red giant branch cannot be explained by photometric errors,
but is due to a spread in metallicity and probably age. Bold dots are definite
cluster members by their radial velocities.
}
\end{figure}

Figure \ref{cmd1} shows the $(b-y)$ versus $V$ color magnitude diagram of all
observed stars in $\omega$ Cen with a photometric error less than 0.1 mag 
($\simeq$ 91450 stars, grey dots) and less than 0.03 mag ($\simeq$ 20620 stars,
black dots) in $y$ and $b$. Furthermore about 300 definite cluster members,
on the basis of their radial velocities (Norris et al. \cite{norr97},
Suntzeff \& Kraft \cite{sunt96}) are marked with bold dots. The colors have
been corrected for a reddening value of $E_{B-V} =
0.11$ mag (Zinn \cite{zinn85}, Webbink \cite{webb}, Reed et al. \cite{reed},
Gonzalez \& Wallerstein \cite{gonz}). This corresponds to $E_{b-y} = 0.08$ and
$E_{m_1} = -0.02$, using the relations of Crawford \& Barnes (\cite{craw}),
$E_{b-y} = 0.7E_{B-V}$ and $E_{m_1} = -0.3E_{b-y}$.
The broad red giant branch (RGB) cannot be explained by photometric errors or
internal reddening. The dispersion in the blue horizontal branch (BHB) along
the reddening vector in the range $-0.01 < (b-y)_0 < 0.05$ and $14.5 < y_ 0 <
14.8$ is 0.037 mag. The expected photometric error in the same direction is
0.017 mag. Thus, internal reddening should be in the order $E_{b-y} \leq 0.02$
mag, or $E_{B-V} \leq 0.03$, consistent with what already was found by
Norris \& Bessell (\cite{norr75}). The dispersion of the RGB in the same
magnitude bin and $0.47 < (b-y)_0 < 0.57$ is 0.051 mag, significantly higher
than that at the BHB. At other magnitudes the dispersion orthogonal to the
RGB is as follows: $12.0 < y_0 < 12.4$: 0.078 (0.013), $13.2 < y_0 < 13.6 $:
0.055 (0.016), and $16.3 < y_0 < 16.7$: 0.035 (0.025).
In brackets the expected photometric errors are given.
Note that the dispersion increases from the lower towards the upper RGB.

\begin{figure}
\psfig{figure=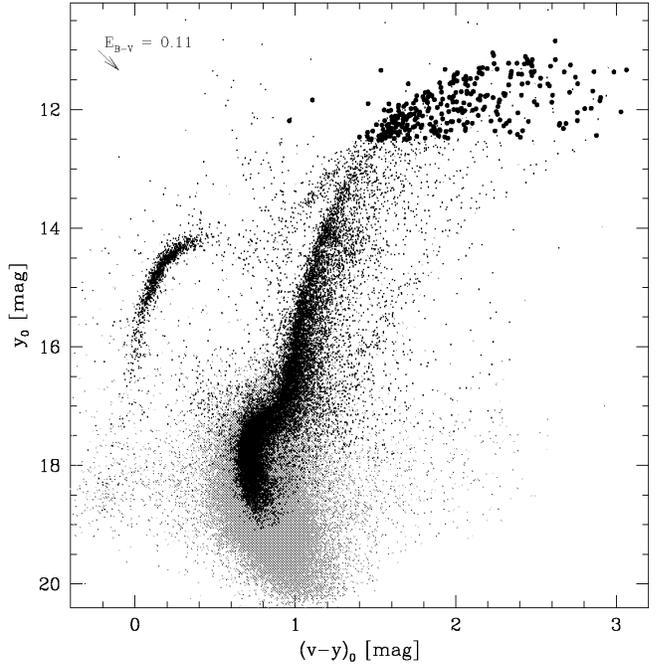,height=8.6cm,width=8.6cm
,bbllx=14mm,bblly=65mm,bburx=188mm,bbury=231mm}
\vspace{0.4cm}
\caption{\label{cmd2} Color magnitude diagram with $(v-y)_0$ as color for the
same stars as in Fig.~\ref{cmd1}. Note the larger scale in color as compared to
Fig.~\ref{cmd1}. The large spread in this CMD is dominated by the CN variations
in the cluster stars. The CN-normal stars of the metal-poor population clearly
define a narrow sequence at the blue side of the RGB
}
\end{figure}

Since the $(b-y)$ color as a metallicity indicator is not affected by CN
variations, the spread in the RGB
must be due to a spread in the overall metallicity (iron abundance)
and/or age. It is interesting to note that a remarkable number
of stars is located redwards of the apparent ``main'' RGB, which cannot be
explained by foreground stars only (a lot of them are definite members, see
Fig.~\ref{cmd1}). Their existence in this CMD points to a
small population of stars that are significantly more metal-rich
than the majority of the stars in $\omega$ Cen (see also Pancino et al. 
\cite{panc}).

The situation changes when plotting the $(v-y)_0$ color versus $y_0$ (see 
Fig.~\ref{cmd2}). The red giants cover a much wider color range. In this CMD 
the scatter towards redder colors is not only due to different metallicities
and ages, but is mainly dominated by the CN variations, since the $v$ filter
is sensitive to the CN band strength. Whereas CN-rich stars scatter to redder
colors, the CN-normal stars with a low metallicity clearly define a sharp
sequence at the blue side of the RGB.
Thus, the combination of both diagrams can be used to separate the metal-poor
CN-normal stars from CN-rich ones (see next section) and study their properties.
Note also that in the $(v-y)$,$y$ CMD the asymptotic giant branch (AGB) is
more clearly separated from the RGB than in the $(b-y)$,$y$ CMD.

\subsection{The redness parameter}

\begin{figure}
\psfig{figure=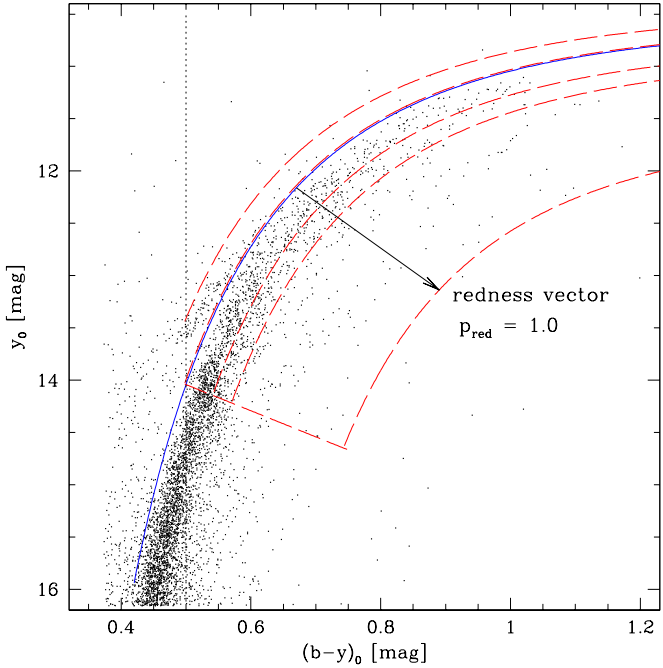,height=8.6cm,width=8.6cm
,bbllx=9mm,bblly=65mm,bburx=195mm,bbury=246mm}
\vspace{0.4cm}
\caption{\label{pred} In this color magnitude diagram all stars with $V_0 <
16.0$ mag, $(b-y)_0 > 0.35$, a photometric error less than 0.05 mag are 
plotted. All stars which are enclosed by the dashed lines were used for the
metallicity determination ($\simeq$ 1500 stars). The solid line indicates
a fit to the blue envelope of the RGB that has been used to define the
zero line for a redness parameter (see text)
}
\end{figure}

As mentioned in the previous section the displacement of giants from
the ``main'' RGB towards redder colors in the $(b-y),y$ CMD reflects their
higher metallicity, but does not depend on the CN anomalies. In order to 
quantify this deviation
we defined a parameter that measures the distance of a red giant
in the CMD from the blue envelope of the RGB along the reddening vector.
The blue envelope was parameterized by  the following function,\\
\centerline{$y_0 = a_0 + a_1/(b-y)_0 + a_2/(b-y)_0^2$}\\
with the parameters $a_0 = 10.917$, $a_1 = -1.308$, and $a_2 = 1.436$.
This in the following called ``redness'' parameter $p_{\rm red}$ is
illustrated in Fig.~\ref{pred}. For illustrative purposes, we distinguish 
between 4 ``populations''. Stars that are located bluewards of
the RGB ($=$ negative $p_{red}$) mainly are AGB stars (P1). The RGB is divided
into three subsamples: a blue (P2; $-0.01 < p_{\rm red} < 0.17$) and a red
(P3; $0.17 < p_{\rm red} < 0.29$) side of the RGB, and all stars that are 
redder than the ``main'' RGB (P4; $0.29 < p_{\rm red} < 1.0$).

%

\section{The two-color diagram and metallicity distribution}

\begin{figure}
\psfig{figure=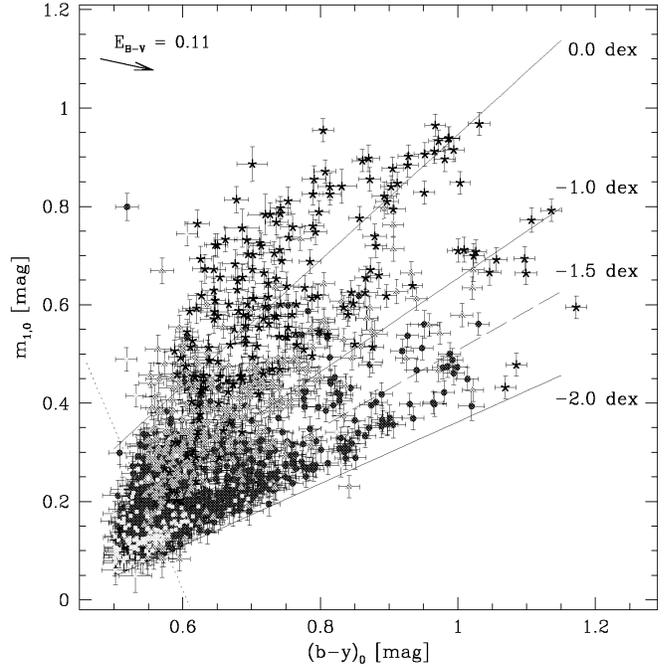,height=8.6cm,width=8.6cm
,bbllx=14mm,bblly=65mm,bburx=188mm,bbury=231mm}
\vspace{0.4cm}
\caption{\label{mplot}
In this plot the $(b-y),m_1$ diagram for about 1500 selected giants (see
Fig.~\ref{pred}) is shown together with the lines of constant metallicity
from the recent calibration by Hilker (\cite{hilk00a}). The error bars include
photometric and calibration errors. The black dots indicate stars from the
blue side of the RGB (P2), grey triangles the ones from the red RGB side (P3),
and asterisks all stars lying apart from the ``main RGB'' (P4). Light grey
squares indicate AGB stars (P1). The dotted line marks the selection criterium
for the metallicity distribution shown in Fig.~\ref{hist1} (lower panel)
}
\end{figure}

\begin{figure}
\psfig{figure=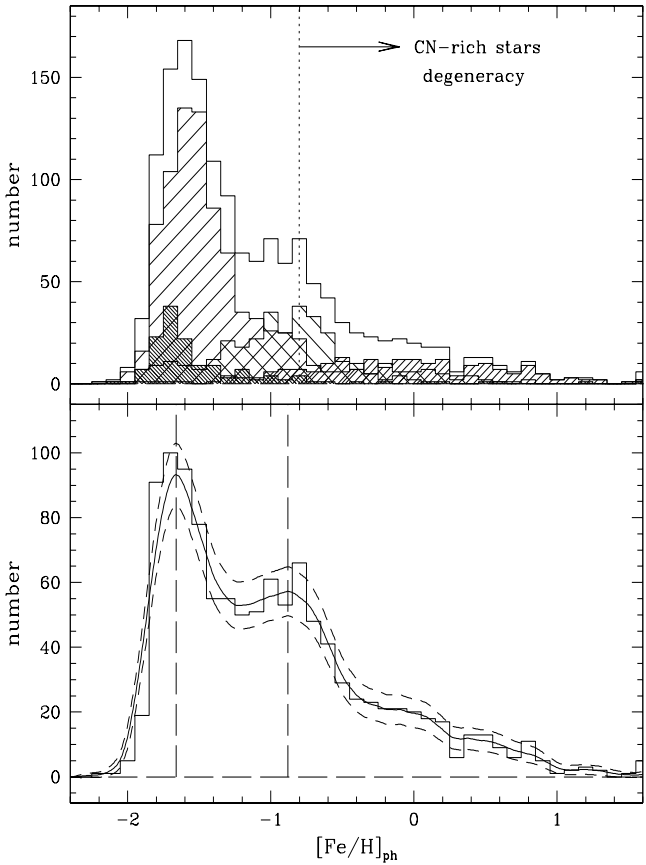,height=11.3cm,width=8.6cm
,bbllx=9mm,bblly=65mm,bburx=159mm,bbury=246mm}
\vspace{0.4cm}
\caption{\label{hist1}
In the upper panel the metallicity distribution of about 1500 red giants is
shown together with the division into subsamples (according to
Fig.~\ref{pred}). Stars from the blue side of the RGB peak around $-$1.7 dex.
AGB stars (more densly hashed) even are more metal-poor. Red RGB stars mainly 
range between
$-1.3$ and $-0.5$ dex, whereas most of the ``reddest'' stars scatter to
higher metallicities. Note that most of the stars with metallicities higher
than $-$0.8 stars are CN-rich stars.
In the lower panel about 1120 stars with a higher accuracy in their
metallicity determination (stars redder than the dotted line in
Fig.~\ref{mplot}) have been selected. The solid and dashed lines represent
the analysis of the metallicity distribution using a density estimation
technique (kernel estimator with an Epanechnikov kernel).
A second peak at about $-$0.9 dex becomes prominent
}
\end{figure}

The $(b-y),m_1$ diagram (Fig.~\ref{mplot}) is indicative for the metallicity
distribution and CN variations of the red giants in $\omega$ Cen.
The selected 1500 red giants (see Fig.~\ref{pred}) show a large scatter
between $-2.0$ and 1.0 dex in their Str\"omgren metallicity.
The Str\"omgren metallicity is defined as
\begin{equation}
{\rm [Fe/H]}_{\rm ph} = \frac{m_{1,0} - 1.277 \cdot (b-y)_0 + 0.331}{0.324
\cdot (b-y)_0 - 0.032}
\end{equation}
following the calibration by Hilker (\cite{hilk00a}).
The trend exists that stars on the blue side of the RGB are mostly
metal-poor, whereas the stars redwards of the ``main'' RGB populate the
metal-rich regime, as one would expect if no CN-anomaly was present.
This trend becomes even more pronounced, if the stars are plotted in a
metallicity histogram. The upper panel in Fig.~\ref{hist1}, where 
[Fe/H]$_{\rm ph}$ denotes the Str\"omgren metallicity,
shows the distribution of all selected stars. A peak around $\simeq -1.7$ dex
with a sharp cutoff towards low metallicities at $-1.9$ dex represents the
blue RGB stars (P2). Also most of the AGB stars (P1) bluewards the main RGB
belong to this metal-poor population.
Stars from the red side of the RGB (P3) have metallicities mainly in
the range $-1.3$ to $-0.5$ dex. If stars with a low accuracy (stars bluer
than the dotted line in Fig.~\ref{mplot}) are skipped, the P3-stars appear
well separated from the metal-poor
population and define a second peak at about $-0.9$ dex (see lower panel,
Fig.~\ref{hist1}), resembling the results of Norris et al. (\cite{norr96}).
Stars with Str\"omgren metallicities higher than about
$-0.8$ dex are supposed to be CN-rich stars of one of the two
populations, since no stars with an iron abundance higher than that has
been found in the cluster.
However, many of them are redder than the ``main'' RGB. Since their $(b-y)$
color is not influenced by CN variations, their existence in the CMD can only
be explained by higher iron abundances.
In the upper panel of Fig.~\ref{hist1} the numbers in the metallicity
histogram represent about the correct proportion of metal-poor to metal-rich
stars, since the stars have been selected due to a luminosity cut in the CMD
(see Fig.~\ref{pred}). The metal-poor population in $\omega$ Cen dominates the
metal-rich one by a ratio of about 3:1.

\subsection{Str\"omgren metallicity versus Fe, C and N abundance}

The metallicity distribution found in our investigation is qualitatively
very similar to that found by Norris et al. (\cite{norr96})
and Suntzeff \& Kraft (\cite{sunt96}) from their Calcium abundance
measurements. In Fig.~\ref{feca} we show the metallicity distribution of those
stars that are in common in both samples. The Calcium abundances from Suntzeff
\& Kraft (upper panel) have been transformed to iron abundances according
to the relation given in their paper.

The behaviour of $\omega$ Cen regarding its relation between Fe abundance
and Str\"omgren metallicity is remarkably different from that of other 
globular clusters (open circles in the lower panel of Fig.~\ref{feca},
including NGC~6334, NGC~3680, NGC~2395, NGC~6397, M22 and M55, taken from
Hilker \cite{hilk00a}).

The straight relation up to -1 dex (with large scatter towards higher 
metallicities) is in striking contrast, for example, to the
situation in M22 (Richter et al. \cite{richp}), where there is a considerable 
scatter at a fixed iron abundance. What determines the Str\"omgren colors 
in $\omega$ Cen?
An answer may come from a comparison of the available elements abundances for
40 giants from Norris \& Da Costa (\cite{norr95}), which in part also entered 
the calibration of Hilker (\cite{hilk00a}) with our Str\"omgren colors. 
Fig.~\ref{fecn} shows in four panels the
Str\"omgren metallicity vs. [Fe/H], [C/H], [N/H], and [C+N/H]. It is apparent
that the correlation with [Fe/H] and [C/H] is very poor. It is better for 
[N/H] (note the large error of 0.4 dex given for the N-abundance by Norris \& 
Da Costa (\cite{norr95}), and best for [C+N/H]. 

\begin{figure}
\psfig{figure=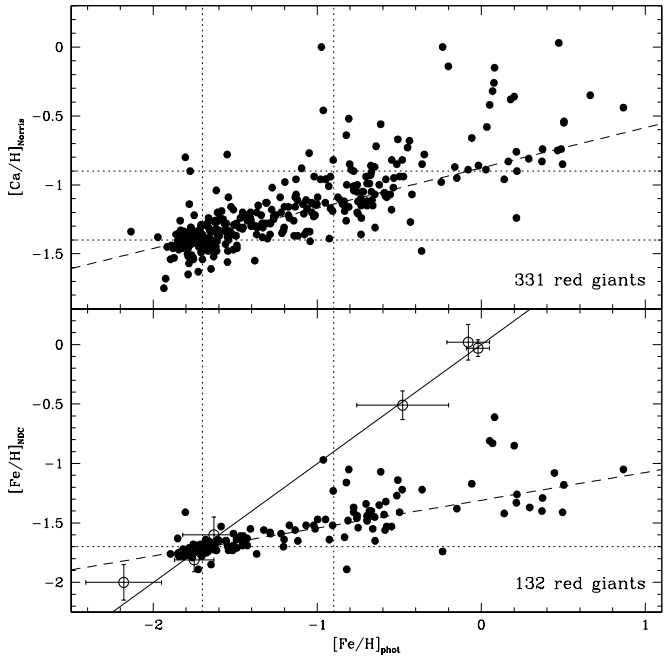,height=8.6cm,width=8.6cm
,bbllx=9mm,bblly=65mm,bburx=195mm,bbury=246mm}
\vspace{0.4cm}
\caption{\label{feca}
The two plots show the metallicity distribution of those red giants that are
in common with the samples of Norris et al. (\cite{norr96}, upper panel) and
Suntzeff \& Kraft (\cite{sunt96}, lower panel).
In the upper panel the calcium abundance is plotted versus [Fe/H]$_{\rm ph}$,
in the lower one the iron abundance as derived from the calcium abundance
using the results of Norris \& Da Costa (\cite{norr95}).
In both panels the dotted lines indicate the detected peaks in the
metallicity distributions. The solid line in the lower panel is the
iso-metallicity line.
Whereas most stars of the metal-poor peak and the mean metallicities of
globular clusters from the literature are located close to this line,
basically all more metal-rich stars clearly deviate from this relation
in a systematic manner (dashed relations).
The CN over-abundance is dominated by the C abundances and seems to be
correlated to the enrichment process of iron.
}
\end{figure}

\begin{figure}
\psfig{figure=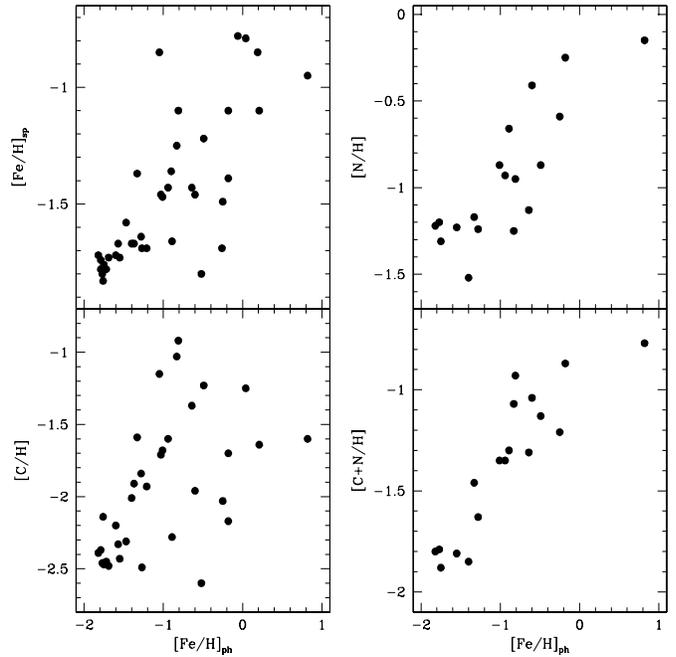,height=8.6cm,width=8.6cm
,bbllx=9mm,bblly=65mm,bburx=195mm,bbury=246mm}
\vspace{0.4cm}
\caption{\label{fecn}
The spectroscopically determined iron, nitrogen, carbon and nitrogen+carbon 
abundances of red giants in $\omega$ Cen (Norris \& Da Costa (\cite{norr95})
are plotted versus their Str\"omgren metallicity. Whereas the correlation 
with [Fe/H] and [C/H] apparently is very poor, it is better for
[N/H] (Norris \& Da Costa give an error of 0.4 dex fot their N-abundances)
and best for [C+N/H]
}
\end{figure}

On the other hand, there is a close correlation of [Fe/H] vs. [C+N/H]
(Fig.~\ref{cpnfe}) (which, by the way, is suprising, given the above large
error of the N abundance). The two most deviating stars are ROA 139 and ROA
144. They have the highest N-abundances in this sample, simultaneously low
oxygen abundances and hence are probably strongly affected by mixing effects.

If we skip them, a linear regression returns $0.64 \pm 0.07$ for the slope, 
indicating that the increase in C+N is faster than in [Fe/H]. Can this be 
understood as a stellar evolutionary effect? As Norris \& Da Costa point out, 
C-depletion as a signature of the CNO-cycle is present and one may see the 
increase in the C-abundance with [Fe/H] in their Fig.8a to have its cause in 
the decreasing efficiency of the mixing-up of processed material with 
increasing metallicity, as it is theoretically expected (e.g. see Kraft 
\cite{kraf94}). But then, the increase
in C+N is not easy to understand, since it is dominated by an increase of N,
where we would expect a decrease, and, after all, the sum of C and N should be
less sensitive to mixing effects.

A further striking fact is that we see in Fig.~\ref{feca} the relation between 
[Fe/H] and
Str\"omgren metallicity already present among the old population, where it is
hard to understand that such small differences in [Fe/H] would cause a regular
pattern in the mixing effects. We thus propose that the gradual enrichment of
C+N, indicated by the Str\"omgren colors, is to a large degree primordial
(we use the term ``primordial'' as the alternative to mixing effects).

The suggestion that a part of the proto-cluster material of $\omega$ Cen has
undergone considerable C-enrichment has also been made by e.g. Cohen \& Bell
(\cite{cohe86}) and Norris \& Da Costa (\cite{norr95}) based on the unique
presence of CO-strong stars.
Moreover, the [C/Fe] abundance in the metal-poor population is about $-$0.7 dex
according to Norris \& Da Costa, which is close to $-$0.5 dex, theoretically
expected from the yield ratios in SNe II (Tsujimoto et al. \cite{tsuj95}). 
If, say, $-$0.2 dex would be the ``true'' value of [C/H], one would require 
a considerable C-contribution
from intermediate-age stars, which by itself is not easy to understand for the
first stars formed in $\omega$ Cen. Also a mean [O/C]-value of approximately
$-$1 dex, as one would read off from Norris \& Da Costa is close to the
theoretically expected yield ratios from SNe II.

\begin{figure}
\psfig{figure=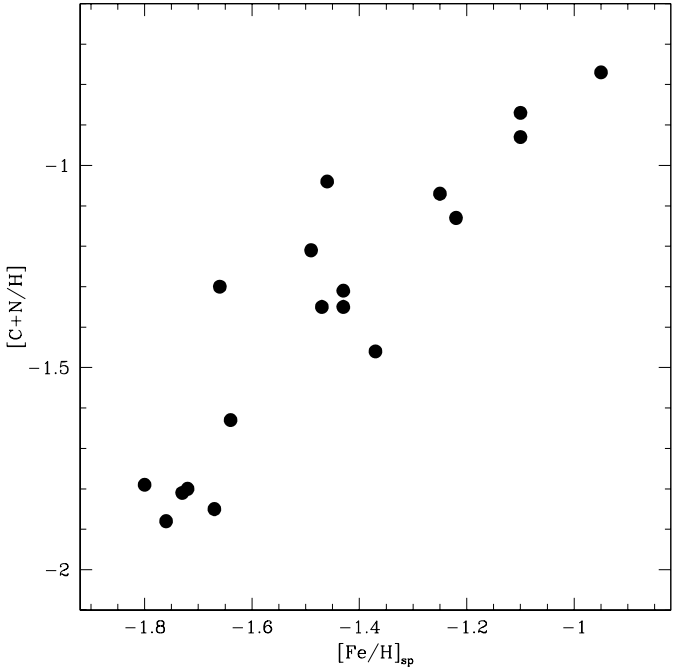,height=8.6cm,width=8.6cm
,bbllx=9mm,bblly=65mm,bburx=195mm,bbury=246mm}
\vspace{0.4cm}
\caption{\label{cpnfe}
The spectroscopically determined iron abundance is plotted versus the sum of
nitrogen+carbon abundances of red giants in $\omega$ Cen (Norris \& Da Costa
(\cite{norr95}). The slope of $0.64 \pm 0.07$ indicates a faster increase in
[C+N/H] than in [Fe/H]. See text for further comments
}
\end{figure}

On the other hand, if the increasing [C+N/Fe] among the metal-poor old 
population was, at least to large part,
primordial, one is driven to the conclusion that the star formation process,
which formed these stars, did not took place in
a single burst within a well-mixed environment, but must have been extended
in time, allowing intermediate-age populations to contribute.

We shall attempt to combine these abundance pattern with other properties
within a consistent scenario later on.

\subsection{Metallicities in the MSTO region}

The Str\"omgren $m_1$ index also is sensible to metallicity for stars around
the main sequence turn-off (MSTO) region. A recent calibration has been
published by Malyuto (\cite{maly}) which is valid in the color ranges
$0.22 < (b-y)_0 < 0.38$ and $0.03 < m_{1,0} < 0.22$ (the above considerations
regarding the C-sensitivity do not necessarily apply here since the CN-band
influence is less in these hotter stars). Hughes \& Wallerstein
(\cite{hugh}) applied this calibration to their Str\"omgren data of $\omega$
Cen and showed that the stars of different metallicity bins hardly differ
in the location at the MSTO region (see their Fig.~9 and 10). There exists
the trend that the more metal-rich stars seem to be bluer in average than the
metal-poor ones, in contrast to what one would expect if all stars had
the same age. Applying Malyuto's calibration to our data we confirm
this finding, as shown in Sect.~5.1. In Fig.~\ref{age} (right panels) the 
location of MSTO stars for three different metallicity bins is shown.
Since our sample is not complete in this magnitude range and photometric
errors are large, we hesitate to attempt a quantitative analysis of the
metallicity distribution in the MSTO region. But it is apparent that the more
metal-rich stars cannot be fit by old isochrones.

\section{Analysis of sub-populations in $\omega$ Cen}

The combination of the redness parameter $p_{\rm red}$ and the Str\"omgren
metallicity [Fe/H]$_{\rm ph}$ can be used to divide the RGB in different
sub-populations and examine their ages
and their spatial distribution in the cluster. In Fig.~\ref{spar}
a plot of $p_{\rm red}$ versus [Fe/H]$_{\rm ph}$ is shown.
As mentioned in the previous section there exists the general trend that the
reddest stars also are tho most metal/CN-rich ones as it is expected.
The solid diagonal line in this diagram shows the expected relation between
$p_{\rm red}$ and [Fe/H]$_{\rm ph}$ if the stars would behave according
to the calibration of Hilker (\cite{hilk00a}) (``CN-normal stars'').
Besides the asymptotic
giant branch, three sub-populations have been selected (boxes with solid 
lines)
that define different parts of the giant branch in the CMD.
Confirmed member stars of $\omega$ Cen according to their radial velocities
are spread all over the parameter space (dark dots). Hence a separation of
non-member stars in this diagram is hardly possible.
Later on, we will use further sub-selections within the sub-populations
(dashed and dotted areas) to  examine the different spatial
distributions within $\omega$ Cen.

\begin{figure}
\psfig{figure=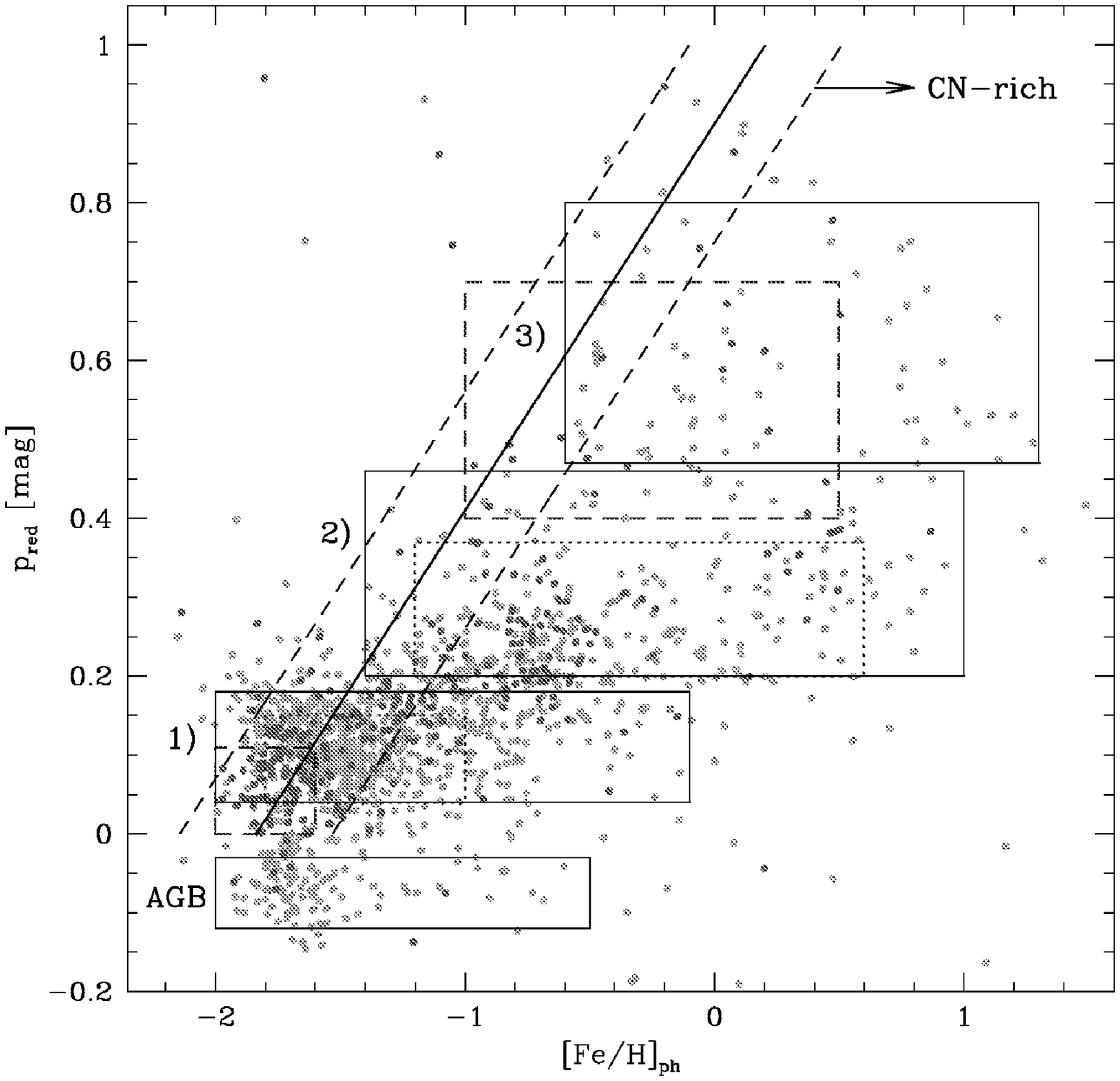,height=8.6cm,width=8.6cm
,bbllx=14mm,bblly=65mm,bburx=188mm,bbury=231mm}
\vspace{0.4cm}
\caption{\label{spar}
The redness parameter $p_{\rm red}$ is plotted versus the Str\"omgren
metallicity. Dark dots mark member stars according to their radial velocities.
The boxes with solid lines define different sub-populations
of the red giants in $\omega$ Cen which have been used for age determination
(see Fig.~\ref{age}). The solid diagonal line shows the expected relation
between $p_{\rm red}$ and [Fe/H]$_{\rm ph}$ for CN-normal stars. The dashed
lines are the error boundaries for $\pm0.3$ dex. Stars within the dotted
boxes in the populations 1) and 2) have been used for the examination of their
cumulative radial distribution (see Fig.~\ref{cum}). In the dashed boxes
stars
are selected for the angular distribution shown in Fig.~\ref{phi}
}
\end{figure}

\subsection{Age estimation of sub-populations}

For the identification of possibly different ages among the sub-populations of
$\omega$ Cen the isochrones from Bergbusch \& VandenBerg (\cite{berg}, in the
following BV92) have been used, converted to Str\"omgren colors by Grebel \&
Roberts (\cite{greb95}).

The location of the metal-poor giants in the CMD has been used to
define a reference isochrone. Adopting their known metallicity ([Fe/H]$= -1.7$
dex), a distance modulus, color offset, and age has been determined. For
fitting the RGB, only stars with a photometric error less than 0.025 mag
in $y$, $b$ and $v$ are selected. The error in metallicity of these stars is
less than 0.3 dex. In the MSTO region stars with a color error less than
0.03 mag in $(b-y)$ and 0.04 mag in $m_1$ have been selected to determine the
location of the isochrone.

\begin{figure*}
\psfig{figure=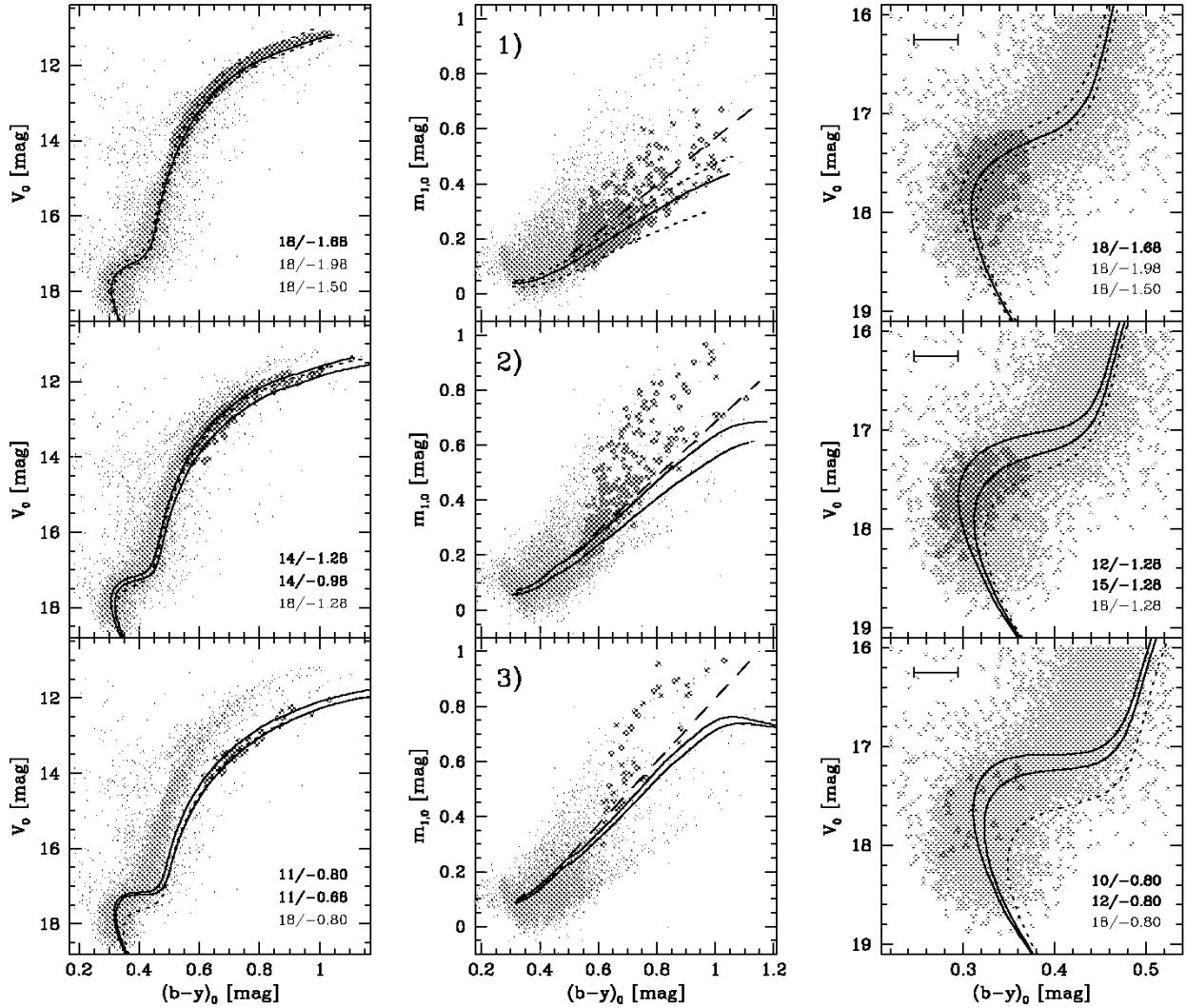,height=15.4cm,width=18.0cm
,bbllx=19mm,bblly=65mm,bburx=199mm,bbury=219mm}
\vspace{0.4cm}
\caption{\label{age}
This figure shows the age and metallicity estimation of three selected
sub-populations (vertical arrangement) in the RGB (dark dots, left panels)
and MSTO region (dark triangles, right panels) of $\omega$ Cen
(see boxes in Fig.~\ref{spar}). Only RGB stars with a
photometric error less than 0.025 mag in $y$, $b$ and $v$ and MSTO stars with
errors less than 0.03 mag in $(b-y)$ were selected (the mean
error bar is indicated in the upper left corner of the right panels).
The error in metallicity of the selected RGB stars is less than 0.3 dex.
The left and right panels show the CMDs of the RGB and MSTO region with
the best fitting isochrones for the corresponding sub-population overlayed.
The isochrones are those from Bergbusch \& VandenBerg (\cite{berg}), converted
to Str\"omgren colors by Grebel \& Roberts (\cite{greb95}).
In the uppermost panels the adopted reference isochrone and two boundary
isochrones (dotted) are shown that fit the MSTO region of the metal-poor stars
and follow the shape of their RGB. In the other panels, the dotted
isochrone has the same age as the reference isochrone, but with a metallicity
that fits the location of the RGB stars in the corresponding selection.
Ages and metallicities of the isochrones are given in the lower right corners
of the plots (thin letters are for the dotted isochrones).
There exists not only an increase in metallicity when going to
redder giant branches, but also a sequence towards younger ages. Isochrones
with high metallicities, but older ages do not fit the MSTO region.
The same isochrones also have been used in the middle panels, where the
$(b-y),m_1$ diagram is shown. The old isochrones cannot be distinguished in
their location from their younger counterparts of the same metallicity.
The redder the giant branch the higher is the
ratio of stars that do not follow the isochrones as determined in the CMD,
but show enhanced CN abundances ($=$ all stars above the dashed lines which
indicates about the separation between CN-normal and CN-strong stars).
The results of the isochrone fitting are summarized in Fig.~\ref{amet}
}
\end{figure*}

The best fitting isochrone for [Fe/H]$=-1.7$ dex is shown in the upper panels
of Fig.~\ref{age} (solid line). A distance modulus of $(m-M)_0 = 13.57$ mag
and additional necessary offsets in $(b-y)$ of $[(b-y)_0-(b-y)_{\rm iso}] = 
-0.037$ mag  and in $m_{1,0}$  of $[m_{1,0}-m_{1,\rm iso}] = 0.005$ were used.
The age that fits best the MSTO region and follows the shape of the RGB
is 18 Gyr. We note that this isochrone does not perfectly fit the color
of the RGB but is slightly redder which might be due to uncertainties in the
color transformations of this set of isochrones to observational data.
Moreover, we are aware of the fact that the absolute age has to be corrected
towards younger ages when applying more recent sets of isochrones (i.e.
VandenBerg et al. \cite{vand00}), but, unfortunately, newly calculated
isochrones in the Str\"omgren system still are missing.

In comparison to the selected reference isochrone, relative ages and
metallicities ($=$ iron abundances) of the other sub-populations in
$\omega$ Cen have been determined (see Fig.~\ref{age}). Whereas the color of
the RGB stars is quite sensitive to metallicity, but not to age, the
metallicity selected stars in the MSTO region can constrain the age
(see also results of Hughes \& Wallerstein \cite{hugh}).

The main (blue) part of the RGB, called population 1) and defined by
$0.04<p_{\rm red}<0.18$ and $-2.0<$[Fe/H]$_{\rm ph}<-0.1$,
corresponds to the reference isochrones. This population can be
fit by isochrones in the metallicity range $-2.0<$[Fe/H]$_{\rm ph}<-1.4$ dex
and ages between 16 and 18 Gyr. Although
all stars in this metallicity range might be fit by the same age (see dotted
lines in the upper panels of Fig.~\ref{age}), there
exists the tendency that the more metal-rich stars seem to be better fit by
the younger isochrones (see Fig.~\ref{amet}).
Stars with higher Str\"omgren metallicities than -1.3 dex (see $(b-y),m_1$
diagram in Fig.~\ref{age}) cannot be fit by
isochrones with the corresponding metallicities. For the same ages these
isochrones would be too red
in the RGB region, for younger ages they are too blue in the MSTO region.
Thus, these stars are most probably CN-rich stars of the old, metal-poor
population.

Population 2), defined by $0.20<p_{\rm red}<0.46$ and
$-1.4<$[Fe/H]$_{\rm ph}<1.0$, lies beyond the main RGB region redwards of
pop 1). Best fitting
isochrones range between $-1.4$ and $-1.0$ dex. Only few stars
show these metallicities in the $(b-y),m_1$ diagram, reflecting the fact that
most of them are CN-strong. The age of this population
ranges between 12 and 16 Gyr. Older isochrones with
metallicities around $-1.3$ dex do not fit the MSTO region (see dotted
isochrone in the right middle panel of Fig.~\ref{age}).
Isochrones with higher metallicities than $-1.0$ dex do not fit pop 2)
at any age.

The last defined population 3), with $0.47<p_{\rm red}<0.80$ and
$-0.6<$[Fe/H]$_{\rm ph}<1.3$, stands apart from the broad main sequence,
but nevertheless contains confirmed cluster members (see Fig.~\ref{spar}) and
was clearly detected in the large sample of Pancino et al. (\cite{panc}).
Isochrones that fit this population are in the range $-0.8$ to $-0.6$ dex
and around 10--12 Gyr. Older isochrones with [Fe/H]$=-0.8$ dex do not fit
the MSTO region.

\begin{figure}
\psfig{figure=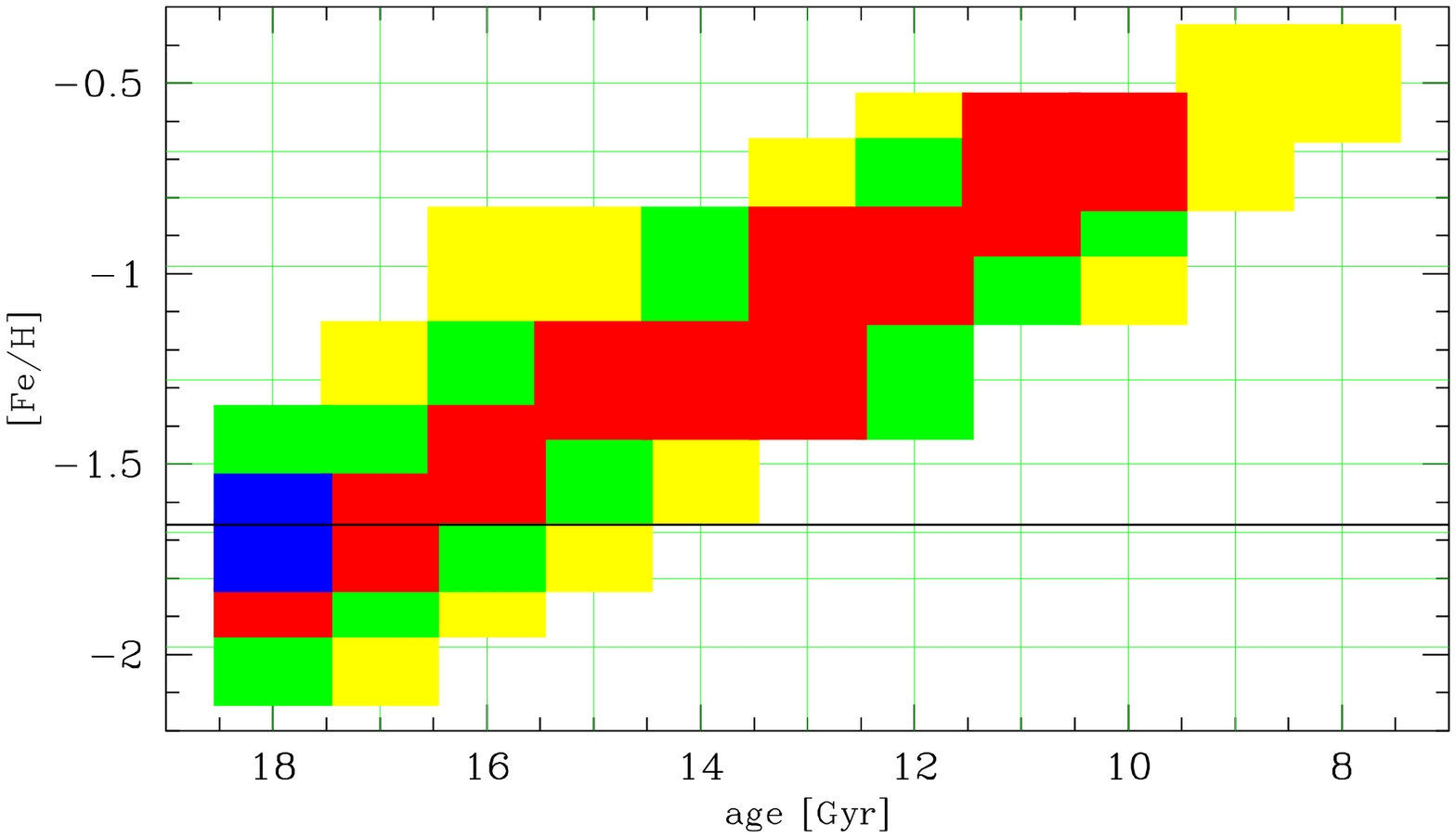,height=4.86cm,width=8.6cm
,bbllx=9mm,bblly=65mm,bburx=195mm,bbury=172mm}
\vspace{0.4cm}
\caption{\label{amet}
In this plot the age-metallicity relation for stellar populations in
$\omega$ Cen is shown. The grid represent the set of isochrones used
(Bergbusch \& VandenBerg \cite{berg}).
The black square defines the adopted reference isochrone.
Dark grey areas are the age and metallicity combinations that fit the CMD 
of $\omega$ Cen best. Middle grey areas are less probable due to their
displacement either in the RGB or the MSTO region,
and light grey are isochrones at the margin of the MSTO region, either
to the blue or to the red side. The horizontal line at $-1.68$ dex marks
the peak of the metallicity distribution (see Fig.~\ref{hist1})
}
\end{figure}

The results of the age determination is summarized in Fig.~\ref{amet}.
In the grid of the isochrone set
dark grey areas mark the best fitting isochrones in comparison to the
reference isochrone (black square). Isochrones represented in
light grey marginally fit the borders of the MSTO and/or RGB region for the
selected stars of a particular metallicity.
There obviously exists an age metallicity relation in the sense that the
more metal-rich stars tend to be younger.
Whereas all stars of the main RGB with metallicities between
$-2.0$ and $-1.4$ dex might be compatible with one age, the populations
with metallicities around $-1.2$ and $-0.7$ dex are at least 2--5 and 5--8 Gyr
younger. If the age metallicity relation in $\omega$ Cen can be understood
as a continuous enrichment process after an initial starburst
with [Fe/H]$\simeq-1.7$ dex, the age spread of the enrichment lies
between 5 and maximally 8 Gyr. We note that the latest isochrones sets
(e.g. VandenBerg et al. \cite{vand00})
give absolute ages that are about 2-5 Gyr younger than that of Bergbusch \&
VandenBerg (\cite{berg}) at a given magnitude and color.
Also the relative age differences would result smaller so that, in fact,
the age spread of the populations in $\omega$ Cen might not be that large 
as indicated above.

\begin{figure*}
\psfig{figure=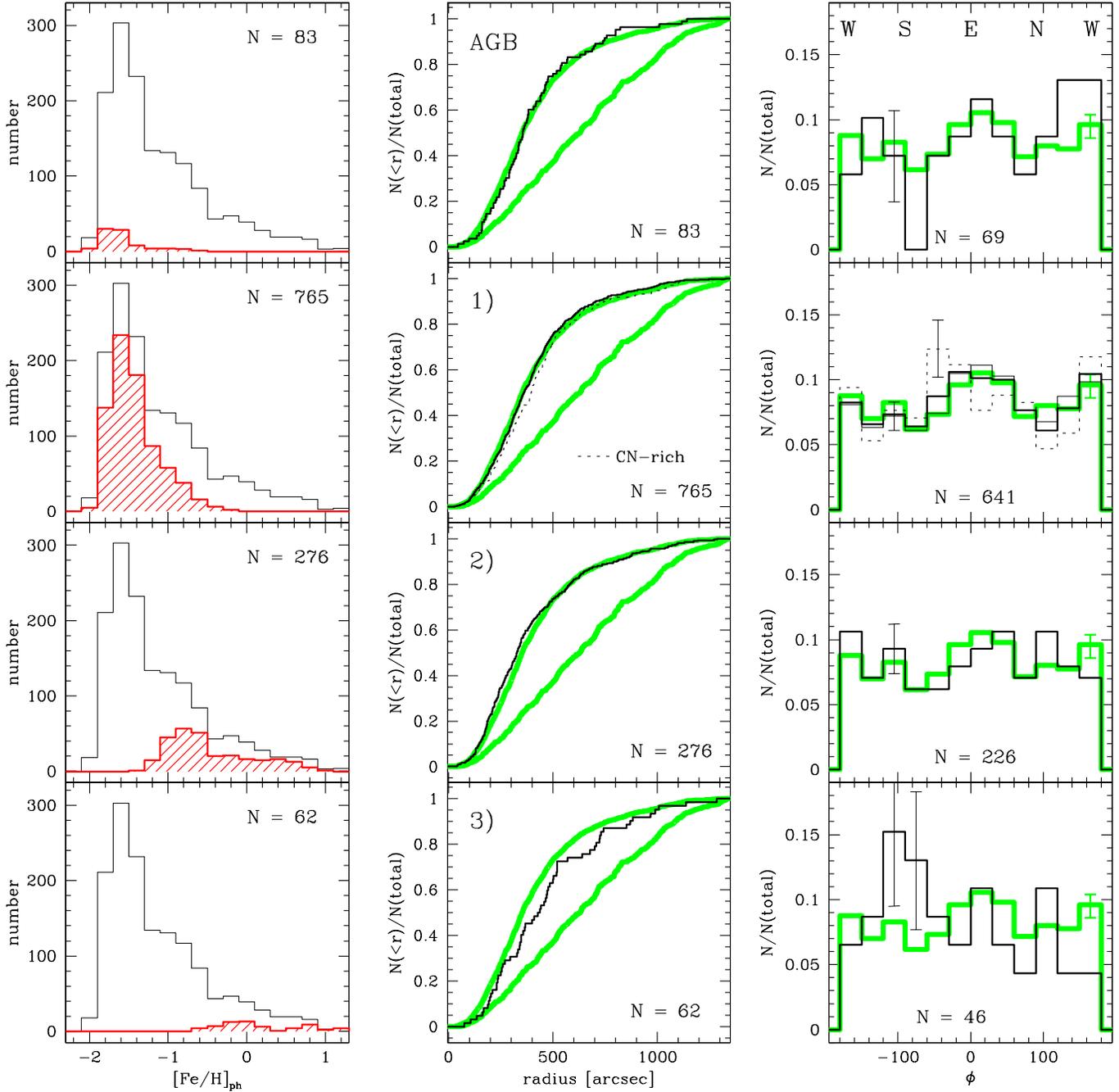,height=18.0cm,width=18.0cm
,bbllx=19mm,bblly=65mm,bburx=199mm,bbury=246mm}
\vspace{0.4cm}
\caption{\label{sub1}
The left panels show the Str\"omgren metallicity distribution of different
sub-populations
of red giants in $\omega$ Cen, selected according to the solid boxes shown in
Fig.~\ref{spar}. As reference the distribution of all giants is shown.
In the middle panels the cumulative radial distributions of the populations
is shown. The grey lines represent reference
distributions of the whole giant sample and probable non-member stars.
Distributions marked with black solid and dashed lines are metallicity
sub-selections as labeled in the plots. The division into CN-normal (thin
solid line) and CN-rich stars (dotted line) in panel 1) is according to the
selection in Fig.~\ref{spar}. The differences of the radial profiles between
the metal-poor stars from pop1 and the CN-rich stars from pop2 and pop3 is
highlighted in Fig.~\ref{cum}.
The angular distribution of the corresponding sample selections is shown
in the right panels. The number counts in the angular bins are normalised to
the total number in each sample. Only stars within a radius of $10\arcmin$
have been selected. The distribution of all selected stars is plotted as
grey histogram for reference. Error bars indicate the average statistical
errors of the corresponding selection in one bin.
In all plots the number of selected stars is given. }
\end{figure*}

\subsection{Spatial distribution of red giants}

The same sub-populations, as defined for the age determination, have been used
to investigate the spatial distribution of their stars. Additionally, the AGB
stars have been included.
In Fig.~\ref{sub1} the metallicity distribution (left panels),
the cumulative radial distribution (middle panels), and the angular
distribution (right panels) of all selections are shown. Only stars within
the area indicated in Fig.~\ref{pred}, with photometric errors less than
0.05 mag and errors in the metallicity determination less than 0.4 dex have
been selected. A central circular area with a radius of
$30\arcsec$ have been excluded because of
incompleteness in this area. The metallicity distribution of all selected
stars ($=1448$) is shown as reference in the left panels.
The cumulative profiles of these stars and 160 probable non-members are shown
as grey lines in the middle panels. The non-members have been selected
according to: $14.5 < y_0 < 16.5$ and $0.1 < (b-y)_0 < 0.4$.
For plotting the angular distribution, we included only stars within a radius
of $10\arcmin$ from the cluster center. The number counts have been normalised
to the total number for each selection. The angle $\phi$ is defined as
$0\degr$
in East direction, $+90\degr$ North, and $-90\degr$ South. The histogram
of all stars ($=1186$ giants, grey histogram in the right panels,
Fig.~\ref{sub1}) shows the reference distribution function in the selected
area.

\begin{figure}
\psfig{figure=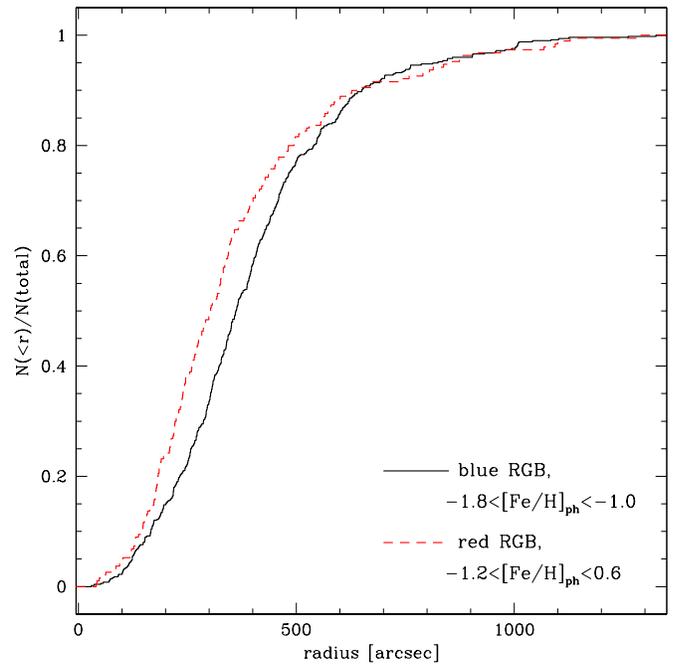,height=8.6cm,width=8.6cm
,bbllx=9mm,bblly=65mm,bburx=195mm,bbury=246mm}
\vspace{0.4cm}
\caption{\label{cum}
This plot shows the cumulative distribution of different subsamples of
red giants in a 20$\arcmin$ broad East-West strip which is perpendicular to the
rotation axis. The selection of the subsamples is illustrated in
Fig.~\ref{spar} (dotted regions). The metal-richer and younger stars (dashed
line) are more concentrated than the old metal-poor population (solid line).
The probability that both populations follow the same radial distribution is
less than 0.1\% (according to a KS test)
}
\end{figure}

The population of the selected 83 AGB stars is very metal-poor with a small
fraction of CN-strong stars (see Fig.~\ref{sub1}, upper left panel).
Therefore, they most probably belong to the old
metal-poor population in $\omega$ Cen. Also their cumulative radial
distributions resemble closely that of pop 1). A Kolmogorov-Smirnov (KS) 
test
gives a 98\% probability that both radial distributions are equal.
Deviations in the angular distribution are statistically not significant as
shown by the error bar.

When comparing population 1) with population 2) there seems to be a slight
difference in their radial distributions in the sense that the ratio
of the metal-rich to metal-poor (pop2:pop1) is higher in the cluster center
than in its outskirts. This result is statistically significant
as discussed below and shown in Fig.~\ref{cum}. The angular distribution
of both sub-populations do not differ significantly from each other.
Within the metal-poor population 1) there exists only a marginal difference
between the 542 selected CN-normal and 209 CN-rich stars in their
spatial distributions.

\begin{figure}
\psfig{figure=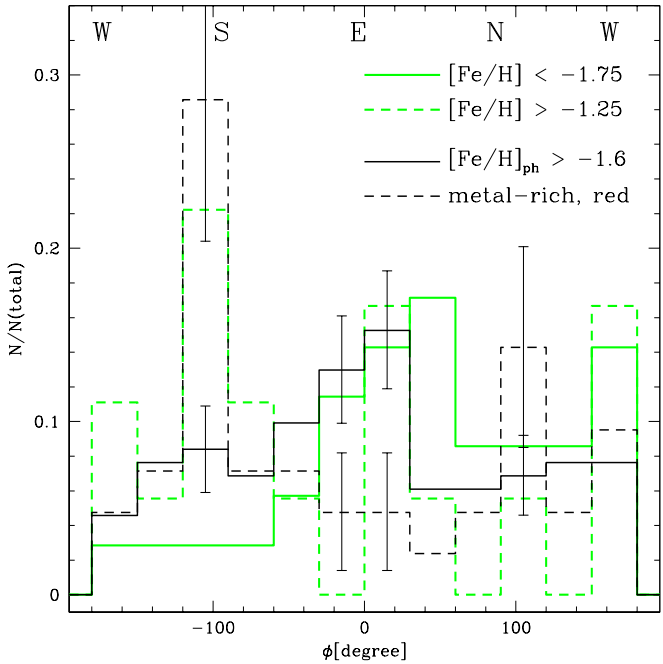,height=8.6cm,width=8.6cm
,bbllx=9mm,bblly=65mm,bburx=195mm,bbury=246mm}
\vspace{0.4cm}
\caption{\label{phi}
The angular distributions of the most metal-poor (solid histograms) and most
metal-rich (dashed histograms) giants in $\omega$ Cen are shown.
The number counts are normalized to the total number in each sample.
The grey distributions represent the samples by Jurcsik (\cite{jurc}).
Black histograms contain stars selected on the basis of our photometric
metallicity determinations (selection areas, see dashed boxes in
Fig.~\ref{spar}).
Errorbars indicate the statistical error in the corresponding bins.
The excess of very metal-rich stars in the South direction could be
confirmed. However, metal-poor stars seem not to be segregated towards the
North, but show the same number counts in South and North direction
}
\end{figure}

In Fig.\ref{cum} the higher concentration of the more metal-rich stars
is highlighted. In this plot only stars in an E-W-strip with a N-S-extension
of 20$\arcmin$ have been selected, since these fields belong to the most
homogeneous set of long exposures. 500 metal-poor giants are compared with
190 giants of the more metal-rich population (selections see dotted boxes in
Fig.~\ref{spar}).
A KS test reveals a probablity of less than 0.1\% that the cumulative number
counts of both populations follow the same radial distribution.

The radial distribution of the stars in pop 3) appears less
concentrated than the average cluster population. Some of them might
be solar metallicity foreground stars. However, others are confirmed cluster
member stars. The angular distribution of pop 3) shows a concentration of
stars towards the South and a slight depression in the West and North
direction which can explain the different radial distribution. In this context
we note that Jurcsik
(\cite{jurc}) reported on a spatial metallicity asymmetry in $\omega$ Cen.
Taking all stars with known iron abundances, either directly measured or
derived from [Ca/H] values, she found that the most metal-rich stars with
[Fe/H]$>-1.25$ dex are concentrated towards the South, whereas the
most metal-poor stars with [Fe/H]$<-1.75$ are more concentrated in the North.
In order to check this result, two sub-samples have been defined that
match the two selections by Jurcsik in our [Fe/H]$_{\rm ph},p_{\rm red}$
plane (dashed boxes in Fig.~\ref{spar}). In Fig.~\ref{phi} the angular
distributions are shown. The sample of Jurcsik (light grey histograms)
contains 35 metal-poor and 18 metal-rich stars that are in common with our
photometric sample. The high concentration of the most metal-rich stars in the
South is confirmed by our selection of 42 metal-rich stars. It even is more
pronounced. Nearly 30\% of the sample are located in the Southern angular bin.
A KS test reveals a probability of 94\% that the metal-rich stars
from the Jurcsik and our sample have the same angular distribution.
However, we cannot confirm an asymmetrical distribution in the North-South
direction of the 131 selected most metal-poor stars.
There exists an excess of stars towards the East which might be explained
by the high ellipticity of this population, but number counts in
the North and in the South are the same. The probability that the angular
distribution of the metal-poor stars of Jurcsik's and our sample agree is less
than 7\% (KS test). Also less than 7\% is the probability that metal-rich and
metal-poor stars are distributed equally.

\section{Problems and a scenario}

Can the younger populations in $\omega$ Cen be enriched by the older one?
In trying to answer this we use oxygen as a tracer for the synthesized
material. First we estimate the number of SNe type II having occured in the 
old population. If we adopt the mass of $\omega$ Cen to be $4\times10^6$ solar 
masses (Pryor \& Meylan \cite{pryor93}), the metal-poor population comprises 
about $2.8\times10^6$ solar masses. Now we assume a Salpeter mass function 
(exponent $-$2.3) between 0.1 and 100 solar masses. Then we can
calculate the number of expected SNe II, if we furthermore assume that each
star more massive than 10 solar masses explodes as SN II. We get 45000 SNe.
This number would be (perhaps considerably) higher, if  
the actual mass function was shallower than a Salpeter mass function
at lower masses, for which there is evidence (Paresce \& De Marchi 
\cite{pare00}). The total oxygen mass
released by these SNe is about 50000 solar masses, (based on Table 7.2 of
Pagel \cite{pagel97}). On the other hand, following Norris \& Da Costa
\cite{norr95}, a mean [O/H]
value for the metal-rich population is -0.7 dex (adopting [O/Fe] = 0.5 dex and 
[Fe/H] = $-$1.2 dex), so we calculate the actual oxygen mass to be 2300
solar masses, if the total mass of the young population is $1.2\times10^6$ 
solar 
masses. Since Smith et al. (\cite{smith00}) see no signature of enrichment by
SNe Ia, it is reasonable to assume that the oxygen mass, which was present
already in the gas before the enrichment, scales with the iron abundance.
The difference is approximately a factor of 3, so we have 1500 solar masses
of newly synthesized oxygen in the metal-rich population. This
means that only 3\% (and probably much less) of the released oxygen has been
retained.  If we do that exercise with the iron abundance, we have 1000 solar
masses of iron released (Pagel \cite{pagel97}, S. 158), and we have about 20 
solar masses of newly synthesised iron present. Of course, the exact numbers 
are insignificant, but they demonstrate that practically {\it all} material 
must have been blown out.

This is also plausible from the energy point of view. We have a release of
kinetic energy by about $5\times10^{54}$ erg from the SNe (neglecting previous 
stellar winds and ionizing radiation), while the binding energy of the
``proto-young population'' is about $2\times10^{52}$ erg, if we for simplicity 
imagine that the gas was confined within an half-light radius of 7 pc 
(Djorgovski \cite{djor93}).

Similar factors must apply for the overall fraction of retained gas, implying
an unreasonable large protocluster mass (neglecting the problem how a bound
system could survive with such low star formation efficiency), if one wants
to live with a permanently retained large gas fraction. 

But we have even more problems. Smith et al. (\cite{smith00}) do detect only 
weak signatures of SNe Ia, expressed by the low [Cu/Fe] value of -0.6 dex. 
On the other hand, both the age spread, the increase of s-process elements, 
and the interpretation of the
Str\"omgren results as primordial enrichment of C and N speak for the 
contribution of an intermediate-age population. Within 3-4 Gyr, at least some 
Ia events should have been ocurred, making the problem with the low iron 
content even worse, if they would have provided iron to the young population. 
Why do we not see their debris? 

Moreover, it is remarkable that we find the signature of intermediate-age
populations already among the old population, indicated by the evidence that
the same relation between [Fe/H] and Str\"omgren metallicity (Fig. 7), which
connects the oldest and the younger population, appears already to be present
at the lowest metallicities. The general behaviour of the Str\"omgren 
metallicities resembles in this respect the well established enrichment of 
s-process elements relative to iron. Unfortunately, the sample of 
spectroscopically studied stars is still too small to allow a definite 
statement also for the metal-poor population, but Fig.12 of Norris \& Da Costa 
(\cite{norr95}) indicates that there might be an increase of barium relative 
to iron already at low metallicities. How can these stars, in a regular 
pattern, be self-enriched simultaneously by SNe II and by intermediate-age 
stars?

It may be that one can construct a scenario in which these oddities can be
explained by pure self-enrichment (see for instance Smith et al. 
\cite{smith00}). However, we wish to point out an alternative, which has not 
yet been mentioned and which seems to offer an easier way towards an 
understanding of $\omega$ Cen.

The above problems arise under one certain assumption, namely that
$\omega$ Cen formed out of a gaseous protocluster of the appropriate mass, 
which had for a long time a high gas-to-star ratio. If we (speculatively) drop 
this assumption, then the problems are less severe.

The previously expressed hypothesis that $\omega$ Cen was the nucleus of a 
dwarf galaxy, gains much attractivity in this context. We additionally 
speculate that its star formation rate was triggered
galaxy. We additionally speculate that its star formation rate was triggered 
over a very extended period (perhaps more than 5 Gyr) by {\it mass supply} 
from the overall gas reservoir of its
host galaxy. This scenario can explain all characteristic properties of
$\omega$ Cen found so far.

This gas inflow, already enriched in the host galaxy, could have occured in
a non-spherical, clumpy and discontinuous manner, providing angular
momentum and thus giving rise to the flattening of $\omega$ Cen. We have no
problems with the competition of gas removal and simultaneous enrichment.
The intermediate-age population stars in $\omega$ Cen released their gas,
for instance by planetary nebulae, in a much less violent fashion and
the infalling gas can mix with this C and N rich material, which also
was rich in s-process elements, giving rise to a new star formation period. 
The large scatter in the Str\"omgren metallicities may thus be in part 
primordial, reflecting the incomplete mixing of the infalling gas with the 
C-N-rich material.
Both SNe II,Ib and Ia would sweep up the gas almost completely, terminating
star formation for a short while, until further mass infall becomes possible. 

It also seems natural that younger and more metal-rich populations show other
kinematic and spatial properties, including asymmetries in their spatial
distribution, depending on the details of the infall process.
We would then expect many periods of strong star formations alternating with
periods of mass infall. The mass infall would finally cease after the gas 
content of the host galaxy has become sufficiently low or was perhaps removed
by ram presure stripping in the Galactic halo during its infall in the
Milky Way (e.g. Blitz \& Robishaw \cite{blit}). One is tempted to ask, whether 
$\omega$ Centauri could have evolved to a object resembling the bulge of a 
spiral galaxy, if its host galaxy would have been more massive.

The subsequent evolution can be sketched as follows: on its retrograde
orbit the dwarf galaxy spiraled towards the Galactic center (Dinescu et al.
\cite{dine99b}). On its way it lose the outermost stellar populations
by tidal stripping, including the two likely member globular clusters
NGC 362 and NGC 6779 (Dinescu et al. \cite{dine99b}). Finally, after
its stellar population dissolved totally, the nucleus $\omega$ Cen
remained and appears now as the most massive cluster of our Milky Way.

A strong motivation for the above scenario comes from the C+N enrichment
relative to iron of the metal-poor population. It would therefore be of extreme
interest to have a larger sample of metal-poor stars (possibly main sequence
stars) spectroscopically analysed to see whether this feature can also be
seen in the s-process elements.
 
\section{Summary and Conclusions}

For about 1500 red giants in $\omega$ Centauri Str\"omgren metallicities
have
been determined. Almost 2/3 of them turn out to be metal-poor, with a peak
in the metallicity distribution at [Fe/H]$_{\rm ph} = -$1.7 dex. This finding
is consistent with previous result by other authors (e.g. Norris et al.
\cite{norr96}, Suntzeff \& Kraft \cite{sunt96}). In the CMD,
this population is best fit by old isochrones (17-18 Gyr:
Bergbusch \& VandenBerg \cite{berg}).
The identifiable AGB stars seem
to belong to this population according to their low metallicities and
small fraction of CN-strong stars. The metallicity range in the main RGB
is about $-1.9<$[Fe/H]$<-1.5$ dex. However all stars of this population are
consistent with a single old age, there might exist an age spread of
maximally 2 Gyr in the sense that the more metal-rich stars are younger.

Beyond the strongly peaked metal-poor population, the metallicity distribution
shows a sharp cutoff towards lower metallicities, but a broad, long tail
towards higher Str\"omgren metallicities. Most of these stars are CN-rich.
Among them, a second major population ($\simeq 1/4$ of the whole population),
peaks around [Fe/H]$_{\rm ph} = -$0.9 dex, defines a giant branch
redwards the main RGB that is best fit by an iron abundance around $-1.2$ dex
and ages 2-5 Gyr younger than that of the oldest population.
Isochrones with older ages do not fit the location of the main sequence
turn-off region.

In the CMD of $\omega$ Cen there exists a red giant branch far redwards
from the main RGB that contains proven cluster member stars according
to their radial velocities. This population exhibits less than
5\% of the whole cluster population. The best fitting isochrones have
an iron abundance around $-0.7$ dex and ages 5-8 Gyr younger than the
oldest population. Str\"omgren metallicities of most of them ($\ga$80\%)
are significantly higher than $-0.7$ dex, identifying them as CN-strong
stars.

The comparison between [Fe/H] abundances derived from high-dispersion 
spectroscopy of Norris \& Da Costa (\cite{norr95}) and 
Str\"omgren metallicities shows a behaviour distinctly different from that
observed in other globular clusters. There is hardly a correlation with
[Fe/H], but a close correlation with [C+N/H].

However, the comparison of Str\"omgren metallicities to the larger sample
of Suntzeff \& Kraft (\cite{sunt96}) shows that there is a coupling to 
the iron abundance indicating that this has a primordial cause. It is
already visible among the metal-poor population and we interpret it as
another manifestation of the well established increasing contribution
of intermediate-age populations with increasing iron abundance.

The comparison of the cumulative radial distribution of the two main
populations in $\omega$ Cen exhibits a higher concentration of the metal-rich
stars within a radius of $10\arcmin$ from the cluster center.
The youngest, most metal-rich population has an asymmetrical distribution
around the cluster center with a concentration towards the South.
Jurcsik (\cite{jurc}) argues that the strong asymmetry of the most metal-rich
stars in $\omega$ Cen points towards a relatively recent effect, since an 
initial ($>15$ Gyr) spatial metallicity anisotropy could not have been 
preserved
that clearly. However, Meylan (\cite{meyl87}) calculated that the relaxation
time of $\omega$ Cen beyond the half-mass radius ($r_h = 2\farcm6$) is 20-30
Gyr, compared to 1 Gyr within the core radius ($r_c = 4\farcm8$). Since most
of the metal-rich stars are located around or beyond the half-mass radius
and furthermore seem to be 5-8 Gyr younger than the old population in
$\omega$ Cen, an initially asymmetric distribution of them is probably still
not relaxed.

Our findings are consistent with a scenario in which enrichment
of the cluster has been taken place over a period of at least 3 Gyr and
maximally 6 Gyr. The conditions for such an enrichment can perhaps be
found in nuclei of dwarf galaxies, where a
secondary star formation is common (Grebel \cite{greb97}). All characteristic
properties of $\omega$ Cen (flattening, abundance pattern, age spread, 
kinematic and spatial differences between metal-poor and metal-rich stars) 
could be understood in the framework of a scenario, where infall of 
previously enriched gas occured in $\omega$ Cen over a long period of time. 
Only the enrichment of nitrogen, carbon, and s-process elements took place 
within $\omega$ Cen, where the infalling gas mixed with the expelled matter 
from AGB stars. 
 
The capture and dissolution of a
nucleated dwarf galaxy by our Milky Way and the survival of $\omega$ Cen
as its nuclues would thus be an attractive explanation for this extraordinary
object. Several recent publications also support this idea and rule out
other possibilities like a chemically diverse parent cloud or a merger
of two clusters (e.g. Majewski et al. \cite{maje99b}, Lee et al. \cite{leey},
Hughes \& Wallerstein \cite{hugh}).

%

\acknowledgements
This research was supported through `Proyecto FONDECYT 3980032'.
We thank Boris Dirsch for helpful discusiions and Eva Grebel for giving 
access to her isochrones.

\enddocument